\documentclass[final,5p,times,twocolumn,12pt]{elsarticle}

\journal{Physics Letters A}

\usepackage{xargs,xifthen,xspace}
\usepackage[utf8]{inputenc}
\usepackage[T1]{fontenc}
\usepackage[english]{babel}
\usepackage{microtype}
\usepackage[intlimits]{amsmath}
\usepackage{amsfonts,amssymb,amsthm}
\usepackage{mathrsfs}
\usepackage[centercolon=true]{mathtools}
\usepackage{braket,faktor,tensor}
\usepackage{accents,suffix,booktabs}
\usepackage{fixltx2e,relsize}
\usepackage{graphicx}
\usepackage{hyperref,bookmark}
\usepackage[english]{varioref}

\newcommand*{\numberset}{\mathbb}

\newcommand*{\Reals}{\ensuremath{\numberset{R}}}

\newcommand*{\im}{\mathrm{i}}

\newcommand*{\Hilbert}[1][H]{\ensuremath{\mathcal{#1}}}

\def\Ketbra#1#2{| #1 \rangle \langle#2 |}

\theoremstyle{definition}

\theoremstyle{plain}

\theoremstyle{remark}

\newcommand*{\tR}{R}
\newcommand*{\tL}{L}
\newcommand*{\tF}{F}

\newcommand\Qca{QCA\xspace}
\newcommand\Qw{QW\xspace}

\usepackage[active]{srcltx} 

\begin{document}

\begin{frontmatter}



\title{Path-integral solution of the one-dimensional Dirac quantum
  cellular automaton}


\author[a,b]{Giacomo Mauro D'Ariano}
\author[a]{Nicola Mosco}
\author[a,b]{Paolo Perinotti}
\author[a]{Alessandro Tosini}
\address[a]{QUIT group, Dipartimento di Fisica, via Bassi 6,
 Pavia, 27100, Italy}
\address[b]{INFN Gruppo IV, Sezione di Pavia, via Bassi 6, 
Pavia, 27100, Italy}

\begin{abstract}
  Quantum cellular automata have been recently considered as a fundamental approach to quantum field
  theory, resorting to a precise automaton, linear in the field, for the Dirac equation in one
  dimension.  In such linear case a quantum automaton is isomorphic to a quantum walk, and a
  convenient formulation can be given in terms of transition matrices, leading to a new kind of
  discrete path integral that we solve analytically in terms of
  Jacobi polynomials versus the arbitrary mass
  parameter.
\end{abstract}


\end{frontmatter}

\section{Introduction}

The simplest example of discrete evolution of physical systems is that of a particle moving on a
lattice.  A (classical) \emph{random walk} is exactly the description of a particle that moves in
discrete time steps and with certain probabilities from one lattice position to the neighboring
positions. This is a special instance of a more general discrete dynamical model known as
\emph{cellular automaton} introduced by von Neumann \cite{neumann1966theory}.  

A quantum version of the random walk, called \emph{quantum walk} (\Qw), was first introduced in
\cite{aharonov1993quantum} where measurements of the $z$-component of a spin-$1/2$ particle decide
whether the particle moves right or left. Later the measurement was replaced by a unitary operator
on the spin-$1/2$ quantum system, also denoted internal degree of freedom or \emph{coin} system,
with the \Qw representing a discrete unitary evolution of a particle state with the internal degree
of freedom given by the spin \cite{ambainis2001one}. In the most general case the internal degree of
freedom at a site $x$ of the lattice can be represented by an Hilbert space $\Hilbert_x$, and the
total Hilbert space of the system is the direct sum of all sites Hilbert spaces. The quantum walk is
also a special case of a \emph{quantum cellular automaton} (\Qca) \cite{grossing1988quantum}, with
cells of quantum systems locally interacting with a finite number of other cells via a unitary
operator describing the single step evolution. {\Qw}s provide the one-step free evolution of
one-particle quantum states, whereas {\Qca}s more generally describe the evolution of an arbitrary
number of particles on the same lattice. However, replacing the quantum state with a quantum field
on the lattice, a \Qw describes a \Qca linear in the field, corresponding to the discrete evolution
of non interacting particles with a given statistics--a ``second quantization'' of the \Qw--which
then can ultimately be regarded as a \Qca.  This is what we call field \Qca in the present paper.

Both {\Qca}s and {\Qw}s have been a subject of investigation in the fields of computer-science and
quantum information, where these notions have been mathematically formalized and studied extensively
in
Refs.~\cite{schumacher2004reversible,arrighi2011unitarity,gross2012index,ambainis2001one,
knight2004propagating,
  ahlbrecht2011asymptotic}.  The interest in this models was also motivated by the use of {\Qw}s in
designing efficient quantum algorithms \cite{ambainis2007quantum, magniez2007quantum,
  farhi2007quantum,childs2003exponential}.

The first attempt to mimic the Feynman path-integral in a discrete physical context is the Feynman
\emph{chessboard problem} \cite{feynman1965quantum} that consists in finding a simple rule to represent
the quantum dynamics of a Dirac particle in $1+1$ dimensions as a discrete path integral.  In
Ref.~\cite{kauffman1996discrete} Kaufmann and Noyes simplify previous approaches
\cite{jacobson1984quantum,karmanov1993derivation} to the Feynman checkerboard, providing a solution
of the finite-difference Dirac equation for a fixed value of the mass. However, such
finite-difference equation has no corresponding \Qw. More recently, following the pioneering papers
\cite{succi1993lattice,meyer1996quantum,bialynicki1994weyl}, a discrete model of dynamics for a
relativistic particle has been considered in a {\Qw}s scenario
\cite{PhysRevA.73.054302,Yepez:2006p4406,darianopla,
  bisio2013dirac,BDTqcaI,d2013derivation,arrighi2013dirac,
  arrighi2013decoupled,farrelly2014causal,farrelly2013discrete}.

In Ref.~\cite{ambainis2001one} Ambainis et al. gave two general ideas for analyzing the evolution of
a walk. One idea consists in studying the walk in the momentum space providing both exact analytical
solutions and approximate solutions in the asymptotic limit of very long time. The other idea is
using the discrete path-integral approach, expressing the \Qw transition amplitude to a given site
as a combinatorial sum over all possible paths leading to that site. Ref.~\cite{ambainis2001one}
provides a path-sum solution of the \emph{Hadamard walk} (the Hadamard unitary is the operator on
the coin system), while Ref.~\cite{konno2002quantum} gives the solution for the \emph{coined} \Qw,
with an arbitrary unitary acting on the coin space.  The same author considered the path-integral
formulation for \emph{disordered} {\Qw}s \cite{konno2005path} where the coin unitary is a varying
function of time.

In this work, we consider the Dirac automaton that has been derived in
Refs.~\cite{BDTqcaI,d2013derivation} from basic principles about the topology of interactions
(unitarity, linearity, locality, homogeneity, isotropy). Such automaton, which is not a coined
\Qw, gives the usual Dirac equation in the relativistic limit of small wave-vectors for lattice step at the Planck
scale. After reviewing the one dimensional Dirac automaton, we solve analytically the automaton in
the position space via a discrete path-integral.  The discrete path corresponds to a sequence of the
automaton transition matrices, which are closed under multiplication. Exploting this feature we derive
the analytical solution for arbitrary initial state.

\section{The one-dimensional Dirac \Qca}

The Dirac \Qca of Refs.~\cite{BDTqcaI,d2013derivation} describes the one-step evolution of a two-component quantum field 
\begin{align*}
  \psi (x,t) :=\begin{pmatrix}
\psi_\tR(x,t)\\\psi_\tL(x,t)\end{pmatrix},\quad (x,t)\in\mathbb{Z}^2,
 \end{align*}
 $\psi_\tR$ and $\psi_\tL$ denoting the {\em right} and the {\em left} mode of the field. Here we
 restrict to one-particle states and the statistics is not relevant, but the presented solution
 could be extended to multi-particle state for any statistics consistent with the evolution. In the
 single-particle Hilbert space $\mathbb{C}^2\otimes l_2(\mathbb{Z})$, we will use the factorized
 basis $\ket{s}\ket{x}$, with $s=\tR,\tL$.

Here the evolution of the field is restricted to be linear, namely
there exists a unitary operator $A$ such that the one step evolution of the
field is given by $\psi(t+1)=U\psi(t) U^\dag=A\psi(t)$.  In the
present case the assumption of locality corresponds to writing
$\psi(x,t+1)$ as
 linear combination of $\psi(x+l,t)$ with
$l=0,\pm1$. Homogeneity here corresponds to $A$ of the
 form 
\begin{align}
A &= A_\tR\otimes T+A_\tL \otimes T^{-1}+A_\tF \otimes I,\nonumber\\
A_\tR &=
	\begin{pmatrix}
		n & 0 \\
		0 & 0
	\end{pmatrix},\;
	A_\tL = 
	\begin{pmatrix}
		0 & 0 \\
		0 & n
	\end{pmatrix},\;
	A_\tF = 
	\begin{pmatrix}
		   0      & \im m \\
		\im m     &    0
	\end{pmatrix},\label{eq:dirac-1d}\\
T &=\sum_{x\in \mathbb {Z}} \Ketbra{x+1}{x},\nonumber
\end{align}
where $A_\tR,A_\tL,A_\tF$ are called \emph{transition matrices}, and $n^2+m^2=1,\;n,m\in\Reals^+$.

\section{Path-sum formulation of the Dirac \Qca}
After $t$ steps one has $\psi(t)= A^t\psi(0)$, and due to linearity the field $\psi(x,t)$ is a
linear combination of the field at the points $(y,0)$ in the past causal cone of $(x,t)$. Each point
$(y,0)$ is connected to $(x,t)$ in $t$ time steps via a number of different discrete paths.
According to Eq.~\eqref{eq:dirac-1d} at each step of the automaton the local field $\psi(y,0)$
undergoes a shift $T^{l}$, $l=0,\pm 1$, with the internal degree of freedom multiplied by the
corresponding transition matrix $A_h$, with $h\in\{\tR,\tL,\tF\}$. A generic path $\sigma$
connecting $x$ to $y$ in $t$ steps is conveniently identified with the string $\sigma=h_t h_{t-1}
\dots h_1$ of transitions, corresponding to the overall transition matrix given by the product
\begin{equation}\label{eq:prod}
\mathcal{A}(\sigma)=A_{h_t}A_{h_{t-1}}\ldots A_{h_1},
\end{equation}
and summing over over all possible paths $\sigma$ and all points $(y,0)$ in the past causal cone of $(x,t)$, one has
\begin{align}
	\label{eq:psi-paths}
	\psi(x,t) =\sum_{y} \sum_{\sigma}
		\mathcal{A}(\sigma)
		  \psi(y,0).
\end{align}
We now evaluate the sum over $\sigma$ in Eq.~\eqref{eq:psi-paths} analytically versus $x,y,t$.
Upon denoting by $r,l,f$ the numbers of $R,L,F$ transitions in $\sigma$, respectively, using 
$t=r+l+f $ and $x-y=r-l$, one has
\begin{align}\label{eq:xr}
 r=\frac{t-f+x-y}{2},\quad
  l=\frac{t-f-x+y}{2}.
\end{align}
The analytical solution is then evaluated observing that the overall transition matrix $\sum_\sigma
\mathcal{A}(\sigma)$ in \eqref{eq:prod} can be efficiently expressed by encoding the transition
matrices as follows
 \begin{align}\label{eq:encoding}
& A_\tR=nA_{00},\quad A_\tL=nA_{11},\quad A_\tF=im (A_{10}+A_{10})\\
&	A_{00}=	\begin{pmatrix}
			1 & 0 \\
			0 & 0
		\end{pmatrix},\quad
	A_{11}=	\begin{pmatrix}
			0 & 0 \\
			0 & 1
		\end{pmatrix},\\
&	A_{10}=	\begin{pmatrix}
			0 & 1 \\
			0 & 0
		\end{pmatrix},\quad
	A_{01}=	\begin{pmatrix}
			0 & 0 \\
			1 & 0
		\end{pmatrix},
\end{align}
with matrices $A_{ab}$ satisfying the composition rule
\begin{align}
A_{ab}A_{cd}=\frac{1+(-1)^{b\oplus c}}{2}A_{ad},
\end{align}
where $\oplus$ denotes the sum modulo 2.
It is now convenient to denote by $\sigma_f=h_t,\ldots h_1$ a generic path having $f$ occurrences of
the $\tF$-transition, and write Eq.~\eqref{eq:psi-paths} as follows
\begin{align}
	\label{eq:psi-paths2}
	\psi(x,t) =\sum_{y}\sum_{f=0}^{t-|x-y|} \sum_{\sigma_f}
		\mathcal{A}(\sigma)
		  \psi(y,0).
\end{align}
In a path $\sigma_f$ the $\tF$ transitions identify $f+1$ slots
\begin{align}\label{eq:slots}
\tau_1\:\tF\:\tau_2\:\tF\:\ldots\ldots\:\tF\:\tau_{f+1},
\end{align}
where $\tau_i$ denotes a (possibly empty) string of $\tR$ and $\tL$.  According to Eq.~\eqref{eq:prod}
the general $\sigma$ cannot contain substrings of the form
\begin{align}\label{eq:cond1}
&h_ih_{i-1}=\tR\tL,\quad h_ih_{i-1}=\tL\tR,\\\label{eq:cond2}
&h_ih_{i-1}h_{i-2}=\tR\tF\tR,\quad h_ih_{i-1}h_{i-2}=\tL\tF\tL,
\end{align}
which give null transition amplitude. Therefore, according to Eq.~\eqref{eq:cond1} each $\tau_i$ in
Eq.~\eqref{eq:slots} is made only of equal letters, i.e. $\tau_i=hh\ldots h$, with $h=R,L$. On the
other hand Eq.~\eqref{eq:cond2} shows that two consecutive strings $\tau_i$ and $\tau_{i+1}$ must be
made with different $h$. This corresponds to having all $\tau_{2i}=hh\ldots h$ and all
$\tau_{2i+1}=h'h'\ldots h'$, with $h\neq h'$. In the following we will denote by $\Omega_R$ and
$\Omega_L$ the sets of strings having $\tau_{2i+1}=RR\ldots R$ and $\tau_{2i+1}=LL\ldots L$,
respectively, for all $i$. 

The above structure for strings $\sigma_f$ can be exploited to determine the matrix
$\mathcal{A}(\sigma_f)$. We consider separately the cases of $f$ even and $f$ odd. For $f$ even one
has
\begin{align}\label{eq:f-even}
 \mathcal{A}(\sigma_f)\!=\!\alpha(f)
 \begin{cases}
A_{00}+A_{11},  & f=t \\
A_{00},         & f<t,\: \sigma_f\in\Omega_R,\\  
A_{11},         & f<t,\: \sigma_f\in\Omega_L,\\
\end{cases}
\end{align}
while for odd $f$ one has
\begin{align}\label{eq:f-odd}
\mathcal{A}(\sigma_f)\!=\!\alpha(f)
 \begin{cases}
A_{10}+A_{01},  & f=t \\
A_{10},         &f<t,\: \sigma_f\in\Omega_R,\\  
A_{01},         &f<t,\:  \sigma_f\in\Omega_L,\\
\end{cases}
\end{align}
with the factor $\alpha(f)$ given by
\begin{align}\label{eq:alpha}
\alpha(f):=(\im m)^f n^{t-f}.
\end{align}

According to Eqs.~\eqref{eq:f-even} and \eqref{eq:f-odd} we can finally restate Eq.~\eqref{eq:psi-paths2} as 
\begin{align}\nonumber
&	\psi(x,t) =\sum_y\sum_{a,b\in\{0,1\}}\sum_{f=0}^{t-|x-y|}
	c_{ab}(f)\alpha(f)	
		A_{ab}\psi(y,0),\\\label{eq:psi-paths3}
\end{align}
where $c_{aa}(2k+1)=c_{01}(2k)=c_{10}(2k)=0$. The coefficients $c_{ab}(f)$ count the
occurrences of the matrices $A_{ab}$ in the transition matrices of all paths $\sigma_f$, and are
given by
\begin{align}\label{eq:coefficients}
 c_{ab}(f)&=\binom{\mu_+- \nu}{\frac{f-1}{2} - 
\nu}\binom{\mu_-+ \nu}{\frac{f-1}{2} + \nu},\\\nonumber
\nu&=\frac{ab-\bar a\bar b}{2},\quad \mu_\pm=\frac{t\pm(x-y)-1}{2},
\end{align}
where $\bar c:=c\oplus1$, and the binomials are null for non integer arguments.
The expression of $c_{ab}$ is computed via combinatorial considerations based on the structure
\eqref{eq:slots} of the paths, and on Eqs.~\eqref{eq:f-even} and \eqref{eq:f-odd}.  Let us start
with coefficients $c_{00}$ and $c_{11}$.  The matrices $A_{00}$ and $A_{11}$ appear only for $f$
even (see Eq.~\eqref{eq:f-even}) in which case one has $\tfrac{f+2}{2}$ odd strings $\tau_{2i+1}$
and $\tfrac{f}{2}$ even strings $\tau_{2i}$.  $A_{00}$ appears whenever $\sigma_f\in\Omega_R$,
namely when the $R$-transitions are arranged in the strings $\tau_{2i+1}$.  This means that we have
to count in how many ways the $r$ identical characters $R$ and $l$ identical characters $L$ can be
arranged in $\tfrac{f+2}{2}$ and $\tfrac{f}{2}$ strings, respectively. These arrangements can be
viewed as combinations with repetitions which give
 \begin{align}\nonumber
  c_{00}(f)=\binom{\frac{f}{2} + r}{r} \binom{\frac{f}{2} + l - 1}{l}=\binom{\frac{t+x-y}{2}}{\frac{f}{2}} \binom{\frac{t-x+y}{2}-1}{\frac{f}{2}-1},
   \end{align}
 where the second equality trivially follows from Eq.~\eqref{eq:xr}.
Similarly $A_{11}$ appears whenever $\sigma_f\in\Omega_L$ which gives
\begin{align}\nonumber
  c_{11}(f)=\binom{\frac{f}{2} + l}{l} \binom{\frac{f}{2} + r - 1}{r}=\binom{\frac{t-x+y}{2}}{\frac{f}{2}} \binom{\frac{t+x-y}{2}-1}{\frac{f}{2}-1}.
   \end{align}
Consider now the other two coefficients $c_{10}$ and $c_{01}$ counting the occurrences of $A_{10}$ and $A_{01}$. The last ones
appears only when $f$ is odd (see Eq.~\eqref{eq:f-even}) and then 
one has the same number $\tfrac{f+1}{2}$ of odd strings $\tau_{2i+1}$ and even strings $\tau_{2i}$.
Counting the combinations with repetitions as in the previous cases we get 
\begin{equation*}
\begin{split}
  c_{10}(f) = c_{01}(f) 
      & = \binom{\frac{f-1}{2} + r}{r} \binom{\frac{f-1}{2} + l}{l} \\
      & = \binom{\frac{t+x-y-1}{2}}{\frac{f-1}{2}}
              \binom{\frac{t-x+y-1}{2}}{\frac{f-1}{2}},
\end{split}
\end{equation*}
which concludes the derivation of the coefficients $c_{ab}(f)$ in Eq.~\eqref{eq:coefficients}. 

The analytical solution of the Dirac automaton can also be expressed in
terms of Jacobi polynomials  $P^{(\zeta,\rho)}_k$ performing the sum over $f$ in
Eq.~\eqref{eq:psi-paths3} 
 which finally  gives
\begin{align}
	\psi(x,t)    & = \sum_y\sum_{a,b\in\{0,1\}}
	                 \gamma_{a,b} 
	                 P^{(1,-t)}_k \left(1 + 2 \left(\frac{m}{n}\right)^2\right)
                     A_{ab} \psi(y,0), \nonumber \\
    k            & = \mu_+ - \frac{a\oplus b + 1}{2}, \nonumber \\
    \gamma_{a,b} & = -(\im^{a\oplus b}) n^t
                     \left(\frac{m}{n}\right)^{2+a\oplus b}
                     \frac{k!\left(\mu_{(-)^{ab}} + 
                         \tfrac{\overline{a\oplus
                             b}}{2}\right)}{(2)_k},
\end{align}
where  $\gamma_{00}=\gamma_{11}=0$ ($\gamma_{10}=\gamma_{01}=0$)
for $t+x-y$ odd (even) and $(x)_k=x(x+1)\cdots (x+k-1)$.


\section{Conclusions}
We studied the one dimensional Dirac automaton, considering a discrete path integral formulation.
The analytical solution of the automaton evolution has been derived, adding a relevant case to the
set of quantum automata solved in one space dimension, including only the coined quantum walk and
the disordered coined quantum walk. The main novelty of this work is the technique used in the
derivation of the analytical solution, based on the closure under multiplication of the automaton
transition matrices. This approach can be extended to automata in higher space dimension.  For
example the transition matrices of the Weyl and Dirac {\Qca}s in $2+1$ and $3+1$ dimension recently
derived in Ref.~\cite{d2013derivation} enjoy the closure feature and their path-sum formulation
could lead to the first analytically solved example in dimension higher than one.

\section*{Acknowledgments}
This work has been supported in part by the Templeton Foundation under
the project ID\# 43796 \emph{A Quantum-Digital Universe}.






\bibliographystyle{elsarticle-num}
\bibliography{bibliography.bib}







\end{document}